\documentclass[english,reprint,aps]{revtex4-1}
\usepackage[T1]{fontenc}
\usepackage[latin9]{inputenc}
\usepackage{color}
\usepackage{amsbsy}
\usepackage{amstext}
\usepackage{amssymb}
\usepackage{graphicx}

\makeatletter

\providecommand{\tabularnewline}{\\}

\makeatother

\usepackage{babel}
\begin{document}
\title{Quantum Interference Theory of Magnetoresistance in Dirac Materials}
\author{Bo Fu}
\affiliation{Department of Physics, The University of Hong Kong, Pokfulam Road,
Hong Kong, China}
\author{Huan-Wen Wang}
\affiliation{Department of Physics, The University of Hong Kong, Pokfulam Road,
Hong Kong, China}
\author{Shun-Qing Shen}
\email{sshen@hku.hk}

\affiliation{Department of Physics, The University of Hong Kong, Pokfulam Road,
Hong Kong, China}
\date{December, 2018}
\begin{abstract}
Magnetoresistance in many samples of Dirac semimetal and topological
insulator displays nonmonotonic behaviors over a wide range of magnetic
field. Here a formula of magnetoconductivity is presented for massless
and massive Dirac fermions in Dirac materials due to quantum interference in scalar impurity scattering potentials. It reveals
a striking crossover from positive to negative magnetoresistivity,
uncovering strong competition between weak localization and weak antilocalization
in multiple Cooperon modes at different chemical potentials, effective
masses and finite temperatures. The work sheds light on the important
role of strong coupling of the conduction and valence bands in the
quantum interference transport in topological nontrivial and trivial
Dirac materials.
\end{abstract}
\maketitle

\paragraph{Introduction}

Topological insulator and semimetal have attracted much attentions
and witnessed impressive theoretical and experimental breakthroughs
in the past decades \citep{hasan2010colloquium,QiXL11rmp,bansil2016colloquium,armitage2018weyl,SQS}.
Recently, an intriguing magnetic-field-driven crossover from positive
to negative magnetoresistance has been widely observed in variety
of topological materials \citep{Kim2013prl,huang2015observation,Zhang16natcomm,Li16natphys,XiongJ15science,zhang2013weak,liang2018experimental,hirschberger2016chiral,zhang2018topological,zhao2016weak,wang2012anomalous},
where a notch-shaped longitudinal magnetoresistance appears in the
vicinity of the zero magnetic field and turns into a negative magnetoresistance
when the magnetic field exceeds some critical value. The origin of
the notch at small field is not completely understood and may arise
from quantum interference effect \citep{CHAKRAVARTY1986193,Kim2013prl,huang2015observation,Li16natphys,Zhang16natcomm}
or the Zeeman energy \citep{XiongJ15science,liang2015ultrahigh}.
The large negative longitudinal magnetoresistance at higher field
is commonly attributed to the chiral anomaly and regarded as a crucial
transport signature for Weyl fermions \citep{Son13prb,hosur2013recent,burkov2014chiral,burkov2015negative,Kim2013prl,huang2015observation,Zhang16natcomm,Lu-17fp}
but some other mechanisms are also proposed \citep{armitage2018weyl,liang2018experimental,dos2016search,arnold2016negative,andreev2018longitudinal,dai2017negative,wang2018intrinsic}.
Furthermore, similar magnetoresistance behaviors have been reported
such as in $\mathrm{Bi}{}_{1-x}\mathrm{Sb}{}_{x}$ \citep{Kim2013prl},
$\mathrm{ZrTe}{}_{5}$ \citep{Li16natphys,mutch2018evidence} and
$\mathrm{Bi_{2}\mathrm{Se_{3}}}$\citep{wang2012anomalous}, which
are near the topological phase transition point and might have nonzero
Dirac mass to mix different chiralities. Such a similarity betokens
that the magnetoresistance in those materials could originate from
the same physical process, obviously, chiral anomaly cannot account
for such a resemblance in the systems with and without well-defined
chirality. As noted, despite of the ongoing scrutiny on the experimental
front, the magnetoresistance near the transition point has stimulated
relatively little theoretical activity.

In this Letter, we have formulated a theory for magnetoresistance
from quantum interference effect in Dirac materials with scalar impurity
potential. Possible contributing Cooperon channels are identified
not only in some limiting regimes but also in the intermediate regime
where some intrinsic symmetries are broken due to variation of the
chemical potential. The existence of the gapless Cooperon channel
of spin-singlet and orbital triplet makes the magnetoresistivity always
positive at small magnetic field, which can also be used to distinguish
different band topology. The competition of multiple Cooperon channels
leads to the non-monotonic magnetotransport behavior in some parameter
ranges. As a demonstration, the formula is applied to analyze the
measured data from a $\mathrm{Cd}_{3}\mathrm{As}_{2}$ sample \citep{zhao2016weak}.
The good agreement of the theoretical fitting suggests that the quantum
interference of Dirac fermions may account for the crossover from
the positive to negative magnetoresistance and its temperature dependence
is dominated by the electron-electron interaction.

\paragraph{Effective Hamiltonian and Method}

The effective Hamiltonian for three-dimensional Dirac materials on
the basis of $|E\uparrow\rangle,|E\downarrow\rangle,|H\uparrow\rangle,|H\downarrow\rangle$
in the framework of the $k\cdot p$ theory can be written as \citep{SQS}

\begin{equation}
H(\mathbf{k})=\hbar v\mathbf{k}\cdot\boldsymbol{\alpha}+m(k)\beta\label{eq:Hamiltonian}
\end{equation}
where $v$ is the effective velocity, $\hbar$ is the reduced Planck
constant, and $\mathbf{k}=(k_{x},k_{y},k_{z})$ is the wave vector.
The Dirac matrices are chosen to be $\boldsymbol{\alpha}=\tau_{x}\otimes(\sigma_{x},\sigma_{y},\sigma_{z})$
and $\beta=\tau_{z}\otimes\sigma_{0}$, where $\boldsymbol{\sigma}$
and $\boldsymbol{\tau}$ are the Pauli matrices acting on spin and
orbital space, respectively. The mass term $m(k)=mv^{2}-b\hbar^{2}k^{2}$
acts as a $k$-dependent effective magnetic field that polarizes the
orbital pseudo-spin $\tau$ along the $z$ direction. The Hamiltonian
(\ref{eq:Hamiltonian}) can be diagonalized by the Foldy\textendash Wouthuysen
transformation \citep{Bjorken-book}: $UH(\mathbf{k})U^{\dagger}=\varepsilon(k)\beta$
with $U=[\varepsilon(k)+\beta H(\mathbf{k})]/\sqrt{2\varepsilon(k)[\varepsilon(k)+m(k)]}$
and $\varepsilon(k)=\sqrt{\hbar^{2}v^{2}k^{2}+m^{2}(k)}$ is the positive
energy spectrum. One can obtain two energy branches $\pm\varepsilon(k)$
and each branch is doubly degenerate owing to the combined time-reversal
and inversion symmetries. The $k$ dependence of the expectation value
of $\tau_{z}$, $\langle\tau_{z}\otimes\sigma_{0}\rangle=m(k)/\varepsilon(k)\equiv\eta(k)$,
reflects the key difference between the bulk electronic structures
of trivial and topological insulators. The topological invariant is
given by $\mathcal{N}=[\mathrm{sgn}(m)+\mathrm{sgn}(b)]/2$, which
is equivalent to the $Z_{2}$ index \citep{Fu-07prb,Shen-12spin}.
Without loss of generality, in the following discussions, we assume
$b$ be negative and the transition between the trivial and nontrivial
topological insulator phases be achieved by changing the sign of $m$.
Besides, for simplicity, we limit the chemical potential $\mu$ to
the positive energy branch, where the degenerate bands have a single
spherical Fermi surface with Fermi radii $k_{f}$ (the positive root
of $\varepsilon(k)=\mu$) and Fermi velocity $v_{f}=\frac{1}{\hbar}\frac{\partial\varepsilon(k)}{\partial k}|_{k_{f}}$.
In the realistic materials, the disorder is always inevitable. Here
we assume a randomly distributed, spin- and orbital-independent scatterers:
$H_{dis}=U(\mathbf{r})1_{4}$ with the correlation function $\langle U(\mathbf{r})U(\mathbf{r}^{\prime})\rangle=\gamma\delta(\mathbf{r}-\mathbf{r}^{\prime}).$

To calculate the conductivity, we employ the Feynman diagram technique
described in Ref. \citep{Note-on-SM}. The conductivity includes two
parts, i.e., the classical conductivity $\sigma_{cl}$ and its correction
$\sigma_{qi}$ due to quantum interference. Here we focus on $\sigma_{qi}$
only. Usually, $\sigma_{qi}$ can be expressed diagrammatically by
a contraction of spin and orbital indices of the Cooperon structure
factor and the Hikami boxes \citep{hikami1980spin,bergmann1984weak}.
The Cooperon structure factor $\Gamma(\mathbf{q})$ can be calculated
by solving the recursive equation: $\Gamma(\mathbf{q})=\Gamma_{0}+\Gamma_{0}\Pi(\mathbf{q})\Gamma(\mathbf{q})$,
where $\Gamma_{0}$ is the bare impurity scattering vertex and $\Pi(\mathbf{q})$
is the single rung of the ladder. One should noted that the \textquotedblright complexity\textquotedblright{}
of $\Gamma$, i.e. its spin and orbital content, originates from the
free Green\textquoteright s functions of the Dirac particles embedded
in the kernel $\Pi(\mathbf{q})$ and not the symmetry of the disorder
correlations $\gamma$. In the present case, the non-diagonal character
of the Green's function leads to $16\times16$ matrix structure for
$\Gamma(\mathbf{q})$. In the spin and orbital singlet-triplet basis
$|C_{\sigma,\sigma_{z}}^{\tau,\tau_{z}}\rangle\equiv\left|\tau,\tau_{z}\right\rangle \otimes\left|s,s_{z}\right\rangle $
{[}$\tau(\sigma)=0,\tau_{z}(\sigma_{z})=0$ represents the pseudo(real)-spin
singlet and $\tau(\sigma)=1,\tau_{z}(\sigma_{z})=\pm1,0$ represent
the pseudo(real)-spin triplets{]}, there are 16 Cooperon modes and
only several effective channels govern the quantum correction to the
conductivity.

\begin{table}
\caption{The ingredients of four effective Cooperon channels $i=0,s,$ and
$t_{\pm}$ in the basis of spin-orbital singlet and triplet states
$|C_{\sigma,\sigma_{z}}^{\tau,\tau_{z}}\rangle\equiv\left|\tau,\tau_{z}\right\rangle \otimes\left|s,s_{z}\right\rangle $,
the effective gaps $z_{i}$ and the weighting factors $\mathcal{F}_{i}$,
which is the trace of product of the Cooperon structure factor $\Gamma(\mathbf{q})$
and the Hikami boxes .}

\begin{tabular}{|c|c|c|c|}
\hline
i  & Cooperon Channel in $|C_{\sigma,\sigma_{z}}^{\tau,\tau_{z}}\rangle$ & $\mathcal{F}_{i}$ & $z_{i}$\tabularnewline
\hline
\hline
$0$ & $|C_{0,0}^{1,1}\rangle,|C_{0,0}^{1,-1}\rangle$ & $\mathcal{F}_{0}=1$ & $z_{0}=0$\tabularnewline
\hline
$s$ & $|C_{0,0}^{1,0}\rangle,|C_{1,0}^{1,1}\rangle,|C_{1,0}^{1,-1}\rangle$ & $\mathcal{F}_{s}$ & $z_{s}$\tabularnewline
\hline
$t_{\pm}$ & $|C_{1,\pm1}^{1,1}\rangle,|C_{1,\pm1}^{1,-1}\rangle,|C_{1,\pm1}^{0,0}\rangle$ & $\mathcal{F}_{t_{\pm}}=\mathcal{F}_{t}$ & $z_{t_{\pm}}=z_{t}$\tabularnewline
\hline
\end{tabular}
\end{table}

\paragraph{The contributing Cooperon channels }

Possible contributing Cooperon modes can be divided into four channels
listed in Table I: the genuine weak antilocalization (WAL) channel
$0$, the doubly degenerate weak localization (WL) channel $t_{\pm}$
and the WL-WAL mixed channel $s$ according to their Cooperon gaps
$z_{i}$. Collecting the contribution of all the ingredients with
same $z_{i}$, we obtain the following expression for each Cooperon
channel:
\begin{equation}
C_{i}(q)=\frac{\mathcal{F}_{i}}{\ensuremath{\ell}_{e}^{2}q^{2}+z_{i}},\label{eq:cooperonmode}
\end{equation}
with $i=0,s,t_{\pm}$. Here, $\ensuremath{\ell}_{e}=\sqrt{\mathcal{D}_{0}\tau}$
is the mean free path with the classical diffusion constant $\mathcal{D}_{0}=v_{f}^{2}\tau/3$
and the elastic relaxation time $\mathcal{\tau}$, and the dimensionless
Cooperon gap $z_{i}$, describing the characteristic length scales
within which particle-hole pairs can propagate without loss. Those
modes with large $z_{i}$ cannot diffuse on long distances and thus
are suppressed. $\mathcal{F}_{i}$ are the summed dimensionless weighting
factors for each channel, specifying how each Cooperon channel contributes
to the conductivity, positive or negative, corresponding to WL or
WAL correction. There are five other modes which do not appear in
Table I since the weighting factor is equal to zero or the Cooperon
gap is very large. The detailed analysis for the ingredients can be
found in Ref. \citep{Note-on-SM}.

\begin{figure}
\includegraphics[width=8.5cm]{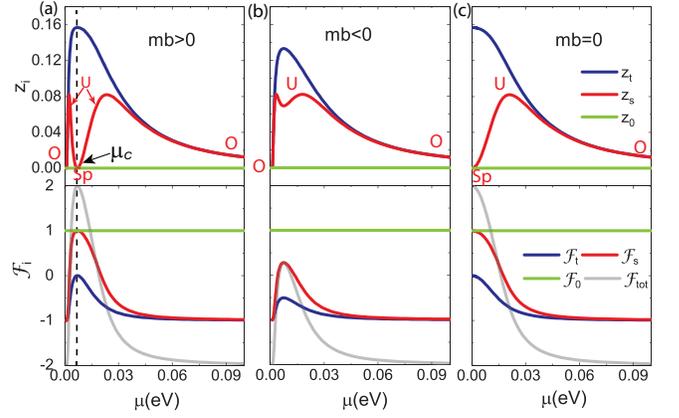}\caption{The dimensionless Cooperon gap $z_{i}$ and the summarized dimensionless
weighting factor $\mathcal{F}_{i}$ as a function of the chemical
potential $\text{\ensuremath{\mu}}$ - for (a) topological insulator
($mb>0$), (b) trivial insulator ($mb<0$) , and (c) Dirac semimetal
($mb=0$). and the mass term(a) $mv^{2}=-0.001\mathrm{eV}$ , (b)
$0.001\mathrm{eV}$ , and (c) $0$ . $\mathcal{F}_{\mathrm{tot}}$
is defined as $\mathcal{F}_{\mathrm{tot}}=\sum_{i}\mathcal{F}_{i}$.
The model parameters are fixed for all calculations in this Letter
to be $b\hbar^{2}=-18\mathrm{eV}\cdot\mathrm{\mathring{A}}^{2}$,
$\hbar v=1\mathrm{eV}\cdot\mathrm{\mathring{A}}$.}
\end{figure}
As shown in Fig. 1, $z_{i}$ and $\mathcal{F}_{i}$ are functions
of chemical potential $\mu$ and exhibit different behaviors for different
band topologies. The behaviors of $z_{i}$ , $\mathcal{F}_{i}$ and
the symmetry pattern for each channel of Dirac semimetal are the same
as topological insulator which begins from $\mu=\mu_{c}\equiv mv^{2}/\sqrt{mb}$,
denoted by the vertical dashed lines in Fig. 1(a). Hence we only need
to discuss the topological trivial and nontrivial cases. Considering
the physical symplectic time-reversal symmetry holds for the full
Hamiltonian regardless of the parameter chosen, there is always one
gapless Cooperon modes with $z_{0}=0$ and $\mathcal{F}_{0}=1$ for
all chemical potentials, which are spin singlet and orbital triplet.
The main difference between the two topological phases is from the
channel $s$ (red lines in Fig. 1). It is a mixture of spin singlet
and triplet and orbital triplet, and there exist a competition between
spin singlet and triplet states. For topological insulator, with increasing
the chemical potential, $z_{s}$ exhibits a multiple crossover: $0\rightarrow\mathrm{finite}\rightarrow0\rightarrow\mathrm{finite}\rightarrow0$
and $\mathcal{F}_{s}$ changes continuously from $-1$ to $1$ and
finally to $-1$. Meanwhile, the symmetry of channel $s$ displays
an evolution as: $\mathrm{O}\rightarrow\mathrm{U}\rightarrow\mathrm{Sp}\rightarrow\mathrm{U}\rightarrow\mathrm{O}$,
where O, U, and Sp represent orthogonal, unitary and symplectic symmetry,
respectively. For trivial insulator, with increasing the chemical
potential, the channel $s$ displays a symmetry pattern: $\mathrm{O}\rightarrow\mathrm{U}\rightarrow\mathrm{O}$,
and $z_{s}$ , $\mathcal{F}_{s}$ vary as $0\rightarrow\mathrm{finite}\rightarrow0$
and $-1\rightarrow1\rightarrow-1$, correspondingly. $z_{t}$ , $\mathcal{F}_{t}$
and the symmetry pattern of the doubly degenerate WL channel $t$
(blue lines in Fig. 1) display similar behaviors for two distinct
topological phases. At band edge and high energy, the Hamiltonian
(\ref{eq:Hamiltonian}) is dominated by $\beta$ term, this channel
becomes gapless ($z_{t}=0$) and $\mathcal{F}_{t}=-1$. In the intermediate
case, we have $z_{t}\ne0$ and $-1<\mathcal{F}_{t}\le0$, channel
$t$ belongs to unitary symmetry class and thus is partially suppressed.

\begin{figure}
\includegraphics[width=8cm]{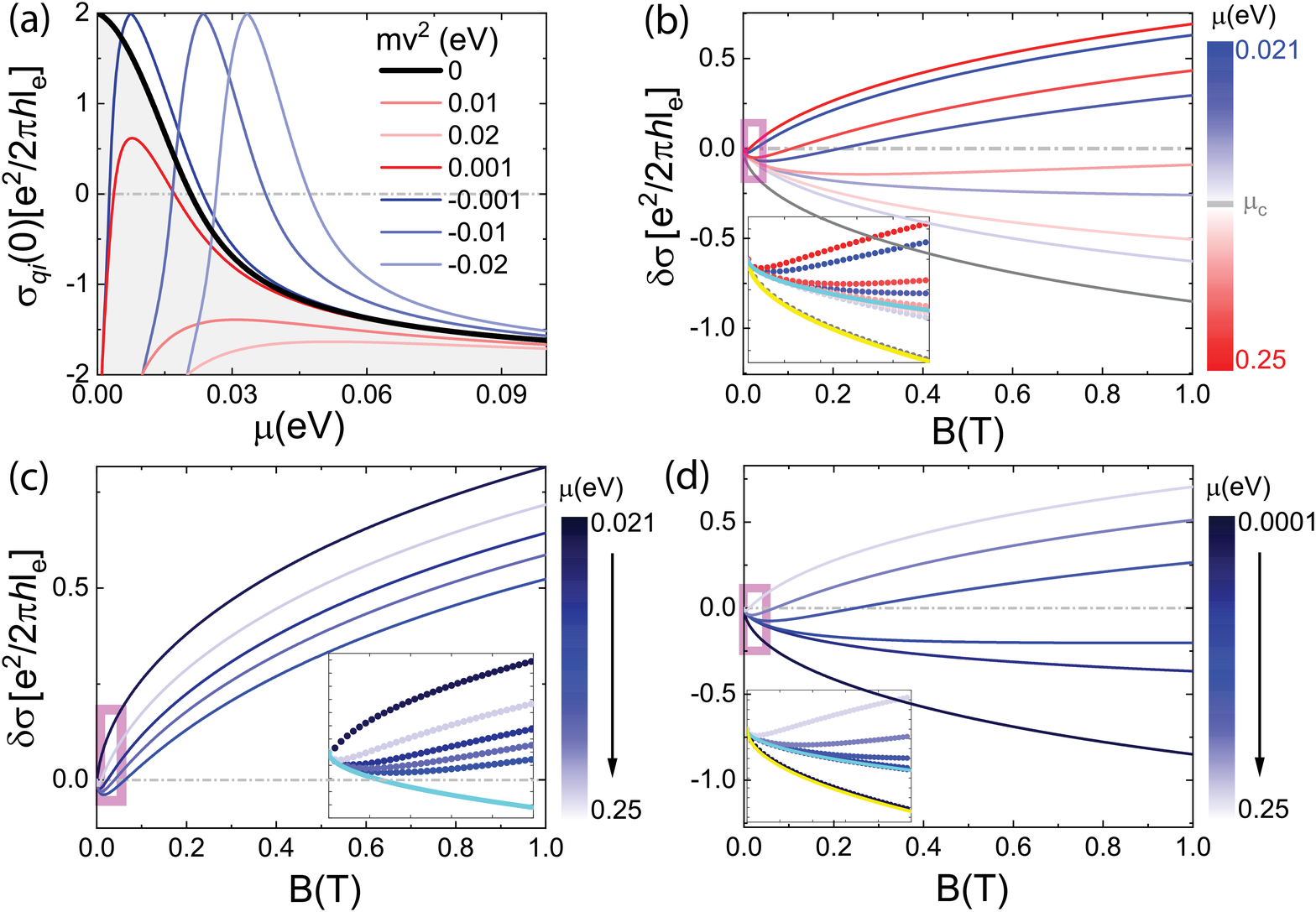}

\caption{(a) The quantum correction to conductivity in unit of $e^{2}/(2\pi h\ell_{e})$
as a function of chemical potential $\mu$ for several different Dirac
mass $m$. The magnetoconductivity in (b) topological nontrivial phase
($mv^{2}=-0.02\mathrm{eV}$), (c) topological trivial phase ($mv^{2}=0.02\mathrm{eV}$)
and (d) Dirac semimetal ($mv^{2}=0$) for several different chemical
potentials. Insets: the enlarged view for small magnetic field denoted
by the pink square region, the green lines and the yellow lines represent
the weak field asymptotic $\alpha\zeta(\frac{1}{2},\frac{1}{2})\frac{e^{2}}{4\pi h\ell_{B}}$
with $\alpha=1$ and $2$ respectively. The coherent length $\ell_{\phi}=100\ell_{e}$
is much larger than the mean free path $\ell_{e}=20\mathrm{nm}$. }
\end{figure}

\paragraph{Conductivity correction from quantum interference }

The conductivity correction from the quantum interference can be obtained
by performing $\mathbf{q}$ integral of Eq. (\ref{eq:cooperonmode})
and summing up the contribution from the four channels in Table I.
Consider the integral over $\mathbf{q}$ diverges in the ultraviolet
limit. The integral usually should be cut off in the ballistic scale.
Similar to Ref.\textcolor{black}{{} \citep{nakamura2018robust}}, we
include the coherent length phenomenologically in the denominator
and introduce a regulating term to make the integral convergent, then
we can find the conductivity correction at the zero magnetic field
($B=0$) as \citep{Note-on-SM}
\begin{equation}
\sigma_{qi}(0)=\frac{e^{2}}{2\pi h\ell_{e}}\sum_{i=0,s,t_{\pm}}\mathcal{F}_{i}\Big(\sqrt{z_{i}+1}-\sqrt{z_{i}+\frac{\ell_{e}^{2}}{\ell_{\phi}^{2}}}\Big)\text{,}\label{eq:quantumcorrectzerofield}
\end{equation}
where $\ell_{\phi}$ is the coherent length due to some inelastic
scattering processes, such as the thermal excitations of atomic lattice
(phonons), and the electron-electron interactions \citep{PALee85rmp}.

Depending on band topology, the quantum corrections show distinct
behaviors as a function of $\mu$. As shown in Fig. 2(a), we plot
$\sigma_{qi}(0)$ {[}in unit of $e^{2}/(2\pi h\ell_{e})]$ as a function
of $\mu$ in different topological phases according to Eq. (\ref{eq:quantumcorrectzerofield}).
For a Dirac semimetals ($m=0$), $\sigma_{qi}(0)$ changes monotonically
from $2$ to $-2$ by increasing $\mu$ (black line) and exhibits
the crossover from WAL ($\sigma_{qi}(0)>0$) to WL ($\sigma_{qi}(0)<0$)
correction. The curve for $m=0$ {[}black solid line in Fig. 2(a){]}
divides Fig. 2(a) into two regions that the trivial phase can only
exist in the shadow region in Fig. 2(a). For topological insulators
(blue lines), $\sigma_{qi}(0)$ initially changes from $-2$ to $2$
as $\mu$ increasing from $|mv^{2}|$ to $\mu_{c}$ and exhibits crossover
from WL to WAL. As $\mu$ further increases, $\sigma_{qi}(0)$ changes
from $2$ to $-2$, exhibiting crossover from WAL to WL. For trivial
insulators (red lines), for small $|mv^{2}|$, $\sigma_{qi}(0)$ displays
a similar $\mu$ dependence as the topological phase except that $\sigma_{qi}(0)$
can not reach up to $2$ due to the suppression of the channel $s$.
This difference in behavior between the trivial and topological phases
is significant for a sizable $|mv^{2}|$. As show in Fig. 2(a), when
$|mv^{2}|=0.01\mathrm{eV}$ and $0.02\mathrm{eV}$, the trivial phase
always exhibits WL correction. As a summary when $\mu=|mv^{2}|$ or
$\infty$, both the topological and trivial phase behave as two copies
of orthogonal class and we will recover the conventional WL case \citep{hikami1980spin,Lu2011prb,garate2012weak,lu2015weak}.
Only in topological phase the whole system behaves as two copies of
symplectic class for $\mu=\mu_{c}$. In the intermediate case, the
channel $s$ can become gapless for $\mu=\mu_{c}$ in topological
phase but not in trivial phase thus always be suppressed.

\paragraph*{Magnetoresistivity}

Experimentally, this WL and WAL effect can be brought out by applying
an external magnetic field. It will induce a decoherence between the
time-reversal trajectories, thus the quantum conductivity correction
is suppressed and gives negative magnetoresistance for WL and positive
magnetoresistance for WAL \citep{hikami1980spin,PALee85rmp}. The
replacement of $\mathbf{q}$ integral in the transverse direction
by an appropriate sum over the effective Landau levels \citep{Note-on-replacement}
gives us the magnetoconductivity as $\delta\sigma(B)=\sigma_{qi}(B)-\sigma_{qi}(0)$
with

\begin{equation}
\sigma_{qi}(B)=\frac{e^{2}}{4\pi h\ell_{B}}\sum_{i=0,s,t_{\pm}}\mathcal{F}_{i}\zeta\Big[\frac{1}{2},\frac{1}{2}+(z_{i}+x^{2})\frac{\ell_{B}^{2}}{\ell_{e}^{2}}\Big]\Big|_{x=1}^{\ell_{e}/\ell_{\phi}},\label{eq:magnetoconductivity}
\end{equation}
here $\zeta(s,t)$ is the Hurwitz zeta function of order $s$ and
argument $z$, and $\ell_{B}=\sqrt{\hbar/4eB}$ is the magnetic length
\citep{Note-on-SM}. The formula is the main result in the present
work, and can be used to fit experimental data.

By using the asymptotic expansion of the Hurwitz $\zeta$ function
\citep{elizalde2008ten}, we find
\begin{equation}
\delta\sigma(B)=\frac{e^{2}}{4\pi h\ell_{B}}\times\left\{ \begin{array}{cc}
\sum_{i}\frac{\mathcal{F}_{i}}{48\ell_{B}^{3}}\Big[\ell_{i}^{3}-\frac{\ell_{e}^{3}}{(z_{i}+1)^{\frac{3}{2}}}\Big], & \frac{1}{\ell_{B}^{2}}\ll\frac{1}{\ell_{0}^{2}}\\
\zeta(\frac{1}{2},\frac{1}{2}), & \frac{1}{\ell_{0}^{2}}\ll\frac{1}{\ell_{B}^{2}}\ll\frac{1}{\ell_{s}^{2}}\\
\mathcal{F}_{tot}\zeta(\frac{1}{2},\frac{1}{2}), & \frac{1}{\ell_{t}^{2}}\ll\frac{1}{\ell_{B}^{2}}\ll\frac{1}{\ell_{e}^{2}}
\end{array}\right.
\end{equation}
where $\ell_{i}=1/\sqrt{\frac{z_{i}}{\ell_{e}^{2}}+\frac{1}{\ell_{\phi_{i}}^{2}}}$
are the effective coherent lengths and $\zeta(\frac{1}{2},\frac{1}{2})\approx-0.605$.
As shown in Figs. 2(b-d), we plot the low-field ($\ell_{B}>\ell_{e}$)
magnetoconductivity with different chemical potential $\mu$ at
extreme low temperature ($\ell_{\phi}\gg\ell_{e}$) for topological
nontrivial insulator, trivial insulator and Dirac semimetal respectively.
Near $B=0$, $\delta\sigma(B)\sim B^{2}$ follows a quadratic dependence
on magnetic field. Since $\ell_{0}=\ell_{\phi}$, at low temperature
this region is sufficient narrow which is invisible in Fig. 2 thus
can be neglected. With increasing $\mu$, we find same magnetoconductivity
behaviors in Dirac semimetal as shown in Fig. 2(d) as the topological
insulator starting from $\mu_{c}$. We only discuss the situations
with finite Dirac mass. The magnetoconductivity behaviors of both
the two distinct topological phases display nonmonotonic $\mu$ dependence.
In the cases of $\mu=|mv^{2}|$ and $\infty$, we recover the conventional
WL case and $\delta\sigma(B)=-2\times\frac{e^{2}}{h}\frac{1}{4\pi}\frac{\zeta(\frac{1}{2},\frac{1}{2})}{\ell_{B}}$.
In the intermediate regime, due to the existence of gapless channel
$0$, both the two phases exhibit $\sqrt{B}$ negative magnetoconductivity
behavior at small magnetic field ($B\ll\frac{\hbar}{4e}\frac{1}{\ell_{t}^{2}}$),
but with different prefactors depending on band topology. In topological
phase as shown in Fig. 2(b), for $\mu\ne\mu_{c}$ ($z_{s}\ne0$),
only the gapless channel $0$ gives a negative magnetoconductivity
being proportional to $\sqrt{B}$ with a universal prefactor independent
of the details of the modeling parameters (green lines in the inset),
while the contributions from all the other channels with finite Cooperon
gap are proportional to $B^{2}$ and can be neglected. For $\mu=\mu_{c}$,
both the channel $0$ and $s$ are gapless and give the same contribution
to magnetoconductivity which is twice the result of $\mu\ne\mu_{c}$
case, i.e.,the yellow line in the inset of Fig. 2(b). However, in
topological trivial case, no matter where $\mu$ locates, the channel
$s$ has a nonzero Cooperon gap and is suppressed, all the magnetoconductivity
curves collapse onto a same universal line {[}see the inset of Fig.
2(c){]}. This remarkable negative magnetoconductivity behaviors at
small field provide us an elegant way to distinguish the different
topological phases through the bulk states transport measurement.
Following this negative magnetoconductivity region at small field,
we find two distinctly different magnetoconductivity behaviors depending
on the sign of $\mathcal{F}_{tot}$ {[}see the grey lines in Fig.
1{]}. For $\mathcal{F}_{tot}<0$ which corresponds to $|(mv^{2}-b\hbar^{2}k_{f}^{2})/\mu|>0.3$,
a crossover from negative to positive magnetoconductivity can be observed.
For $\mathcal{F}_{tot}>0$, the magnetoconductivity decreases monotonically
as a function of the magnetic field. No positive magnetoconductivity
is observed.

\begin{figure}[!t]
\includegraphics[width=8cm]{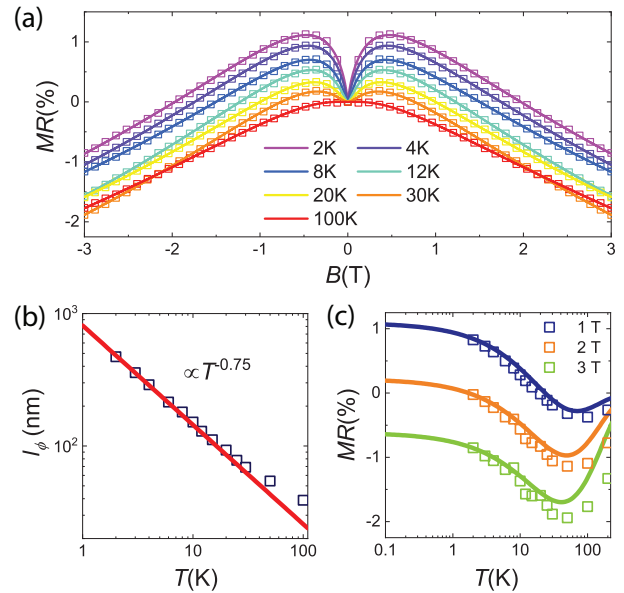} \caption{(Color online). (a) The relative longitudinal magnetoresistance (MR)
of a $\mathrm{Cd}_{2}\mathrm{As}_{3}$ sample. The measured data (open
squares) are extracted from Fig. 2(b) in Ref. \citep{zhao2016weak}
and the solid lines are fitted by using Eq.\ref{eq:magnetoconductivity}
at different temperature $T$. (b) The temperature dependence of the
fitted coherent length $\ell_{\phi}$ (open squares) . The red straight
line indicates the temperature dependent coherent length $\ell_{\phi}\propto T^{-0.75}$
arising from electron-electron interaction as predicted in Ref. \citep{PALee85rmp}.
(c) The relative magnetoresistance as a function of temperature at
several magnetic field strengths. The calculation parameters for the
solid lines are $\rho_{0}=20.2$ $\mathrm{m}\Omega\cdot\mathrm{cm}$,
$\ell_{e}=8.5\mathrm{nm}$, $\eta^{2}=0.268,$ and $\ell_{\phi}=472(T/2\mathrm{K})^{-3/4}\mathrm{nm}$.}
\end{figure}
As a demonstration, we apply the present theory to analyze the measured
longitudinal magnetoresistance of a $\mathrm{Cd_{3}As_{2}}$ sample
in Ref. \citep{zhao2016weak}. $\mathrm{Cd_{3}As_{2}}$ is a three-dimensional
Dirac semimetals, and has been studied extensively \citep{WangZ-13prb,LiuZK-14nm,Neupane-14nc,Jeon-14nm,LiH-16nc}.
In most experiments, a crossover from positive to negative magnetoresistance
has been clearly observed under the longitudinal configuration. A
sharp dip is gradually weaken with increasing temperature. To compare
with the experimental data, we convert the magnetoconductivity $\delta\sigma(B)$
in Eq.(\ref{eq:magnetoconductivity}) into the relative magnetoresistance
as $\mathrm{MR}=-\delta\sigma(B)/[\rho_{0}^{-1}+\delta\sigma(B)]$
with $\rho_{0}$ being the experimentally measured resistivity at
$B=0$. Fig. 3(a) shows an excellent agreement between the fitting
curves (solid lines) and experimental data (open squares) at different
temperatures. It is found that the extracted coherent length $\ell_{\phi}$
fits well with the temperature dependence $\ell_{\phi}\propto T^{-3/4}$
at low temperature as shown in Fig. 3(b), which implies that the decoherent
mechanism is dominated by the electron-electron interactions \citep{PALee85rmp,efros1985electron}.
For fixed $B$, the relative magnetoresistance displays anomalous
nonmonotonic temperature dependence due to the competition of multiple
Cooperon channels in Fig. 3(c). This nonmonotonic behavior disappears
for the system with single WL or WAL correction or at a weak field
\citep{lu2015weak,velkov2018transport}. Different from the two-dimensional
WL and WAL \citep{hikami1980spin,tkachov2011weak,shan2012spin,lu2014finite},
the relative magnetoresistivity in three dimensions saturates at extremely
low temperature ($\ell_{\phi}\to\infty$) \citep{PALee85rmp}. All
fitting parameters in $\delta\sigma(B)$ (see in Ref.\citep{Note-on-SM})
look reasonable and self-consistent. Thus, the good agreement between
the experimental data and theory demonstrates that the crossover of
magnetoresistance is attributed to the quantum interference of Dirac
fermions in the Dirac semimetal.

\paragraph{Discussion and conclusion}

The nonzero mass term in the Dirac Hamiltonian couples Weyl fermions
with opposite chirality, hence the spin and pseudo-spin degrees of
freedom are highly entangled for Dirac materials. To capture the effect
of strong spin-orbit entanglement correctly, it is necessary to treat
all the possible contributing Cooperon modes on the same footing,
which requires one to retain the matrix structure of all the Green\textquoteright s
functions \citep{adroguer2012diffusion,araki2014weak}. The variation
of chemical potential controls the coupling strength between the conduction
and valence bands and causes the interplay of different Cooperon modes.
Thus, all peculiarities of the system are rooted in the spinor-like
character of carrier wave functions rather than the symmetry of the
disorder correlations. The inclusion of other types of disorder \citep{lu2015weak,liu2017weak,adroguer2015conductivity,liu2017quantum}
in our calculations may further break the corresponding time reversal
symmetry of each Cooperon channel and introduce an additional Cooperon
gaps being proportional to disorder strength suppressing its contribution
\citep{ostrovsky2012symmetries,gornyi2014quantum}.

In summary, we have developed a theory for magnetoresistance from
quantum interference effect in Dirac materials with scalar impurity
potential by means of Feynman diagrammatic technique. Possible contributing
Cooperon channels are identified not only in some limiting regimes
but also in the intermediate regime where some intrinsic symmetries
are broken due to variation of the chemical potential. The competition
of multiple Cooperon channels leads to the nonmonotonic magnetotransport
behavior. A finite magnetic field tends to suppress WAL and to release
WL from spin- and/or orbital triplet Cooperon before destroying the
quantum interference completely, uncovering a crossover from a positive
to negative magnetoresistivity. Our finding shows that this crossover
is a consequence of quantum interference of Dirac fermions for a large
class of Dirac materials with a strong coupling of the conduction
and valence bands.

We would like to thank Fengqi Song and Baigen Wang for providing original
experimental data in Fig. 3. This work was supported by the Research
Grants Council, University Grants Committee, Hong Kong under Grant
No. 17301717.

\bibliographystyle{apsrev}

\begin{thebibliography}{localization}
\bibitem{Note-on-replacement} The replacement works very well for the longitudinal magnetoresistance. For the transverse case, the
anisotropic effect of magnetoresistance is suppressed for
a finite field in this approximation.
\end{thebibliography}

\end{document}